\documentclass[reprint,amsmath,amssymb,pre,twocolumn]{revtex4-2}
\usepackage[utf8]{inputenc}

\usepackage{bm}
\usepackage{graphicx}
\usepackage{upgreek}
\usepackage{amsmath}
\usepackage{amssymb}
\usepackage{numprint}
\usepackage{xspace}
\usepackage{comment}
\usepackage[normalem]{ulem}
\usepackage[dvipsnames]{xcolor}

\setcitestyle{numbers,square}

\definecolor{myred}{rgb}{0.78, 0, 0}
\definecolor{myblue}{rgb}{0, 0, 0.78}

\usepackage{hyperref}
\hypersetup{colorlinks = true, citecolor = myblue, linkcolor = myblue}

\newcommand{\diff}{\mathrm{d}}
\newcommand{\vect}{\boldsymbol}

\newcommand{\Eqref}[1]{Eq.~(\ref{#1})}
\newcommand{\figref}[1]{Fig.~\ref{#1}}

\renewcommand{\Re}{\operatorname{Re}}
\renewcommand{\Im}{\operatorname{Im}}
\newcommand{\chiA}{\chi_\mathrm{A}}
\newcommand{\chiR}{\chi_\mathrm{R}}
\newcommand{\chiAideal}{\chi_\mathrm{A}^{\mathrm{eq}}}
\newcommand{\chiRideal}{\chi_\mathrm{R}^{\mathrm{eq}}}

\newcommand{\br}{\mathbf{r}}
\newcommand{\sgn}{\text{sgn}}
\newcommand{\CGLE}{\text{CGLE}}
\newcommand{\EVr}{\lambda_\mathrm{r}}
\newcommand{\qr}{q_\mathrm{r}}
\newcommand{\qrm}{\qr^\mathrm{m}}
\newcommand{\lrm}{l_\mathrm{r}^\mathrm{m}}
\newcommand{\EVc}{\lambda_\mathrm{c}}
\newcommand{\qc}{q_\mathrm{c}}
\newcommand{\qcm}{\qc^\mathrm{m}}
\newcommand{\lc}{l_\mathrm{c}}
\newcommand{\lcm}{\lc^\mathrm{m}}
\newcommand{\lCorr}{l_\mathrm{corr}}

\newcommand{\nocontentsline}[3]{}
\newcommand{\tocless}[2]{\bgroup\let\addcontentsline=\nocontentsline#1{#2}\egroup}

\begin{document}

\title{Influence of physical interactions on spatiotemporal patterns}

\author{Chengjie Luo%
  \email{chengjie.luo@ds.mpg.de}}

\author{David Zwicker%
  \email{david.zwicker@ds.mpg.de}}
\affiliation{Max Planck Institute for Dynamics and Self-Organization, Am Faßberg 17,
  37077 Göttingen, Germany}

\begin{abstract}
Spatiotemporal patterns are often modeled using reaction-diffusion equations, which combine complex reactions between constituents with ideal diffusive motion. Such descriptions neglect physical interactions between constituents, which might affect resulting patterns. To overcome this, we study how physical interactions affect cyclic dominant reactions, like the seminal rock-paper-scissors game, which exhibits spiral waves for ideal diffusion. Generalizing diffusion to incorporate physical interactions, we find that weak interactions change the length- and time-scales of spiral waves, consistent with a mapping to the complex Ginzburg-Landau equation. In contrast, strong repulsive interactions typically generate oscillating lattices, and strong attraction leads to an interplay of phase separation and chemical oscillations, like droplets co-locating with cores of spiral waves. Our work suggests that physical interactions are relevant for forming spatiotemporal patterns in nature, and it might shed light on how biodiversity is maintained in ecological settings.

\end{abstract}

\maketitle
\tableofcontents

\section{Introduction}

Complex spatiotemporal patterns are ubiquitous in nature.
Examples on microscopic scales include the Belousov-Zhabotinsky (BZ) reaction \cite{zaikin1970concentration}, chemical waves created by amoebae~\cite{palsson1996origin}, and electrical patterns in human hearts~\cite{davidenko1992stationary}.
On larger scales, complex patterns emerge in bacterial colonies \cite{hibbing2010bacterial,nadell2016spatial}, lizard populations \cite{corl2010selective}, and  human society \cite{semmann2003volunteering,wang2014social}.
In all cases, patterns emerge from spatial motion and local interactions, like chemical reactions, mating, and competition.
These dynamics are typically modeled as reaction-diffusion equations, where non-linear reactions are combined with ideal diffusive motion~\cite{Cross2009}.
This choice, however, implies that physical interactions that give rise to non-linear local behavior are neglected in the spatial dynamics.
To fill this gap, we here investigate the role of physical interactions on a typical model of spatiotemporal patterns.

Cyclic dominant interactions, like the seminar rock-paper-scissors game~\cite{zhou2016rock,szolnoki2014cyclic,szolnoki2020pattern}, naturally produce temporal oscillations~\cite{reichenbach2006coexistence}.  %
Combined with ideal diffusion~\cite{reichenbach2007mobility} or hopping~\cite{szczesny2013does,szczesny2014characterization}, cyclic dominant reactions produce spatio-temporal patterns.
In particular, spiral waves form when the mobilities of species are low, while spatial patterns are lost for large mobilities~\cite{reichenbach2007mobility, mobilia2016influence}.
Spatial patterns also often subside when random mutations are too prevalent~\cite{nagatani2020diffusively,szczesny2013does,szczesny2014characterization,mobilia2010oscillatory}.
Interestingly, many of these models can be reduced to the complex Ginzburg-Landau equation (CGLE), e.g., by projection onto a reactive manifold \cite{frey2010evolutionary} or a  multiscale expansion \cite{szczesny2013does}.
Such mappings allow to determine parameter regions of spatiotemporal patterns, including vortices, spiral waves, and spatiotemporal chaos~\cite{aranson2002world, van1999sources, aranson1993theory}.

In this paper, we consider a general model of cyclic dominant reactions coupled to diffusive motion including physical interactions.
In the absence of reactions, the physical interactions can lead to phase separation, where all species co-segregate from the inert solvent (for strong attraction) or all segregate from each other (for strong repulsion).
We recently analyzed the effect of such interactions on static Turing patterns and found that even weak interactions, which would not lead to phase separation by themselves, can strongly affect the resulting patterns~\cite{menou2023physical}.
While we here identify similar behavior for cyclic dependent reactions, we also discover entirely new spatiotemporal patterns for strong interactions.
To introduce all these effects in detail, the paper is organized as follows: We introduce the model in section~\ref{sec:model}, identify six relevant parameter regions using linear stability analysis in section~\ref{sec:lsa}, and then discuss these regions in detail using numerical simulations and more detailed analysis in the subsequent sections.

\section{Results}
\subsection{Model with physical and chemical interactions}
\label{sec:model}

\begin{figure*}
  \centering
  \includegraphics[width=\textwidth]{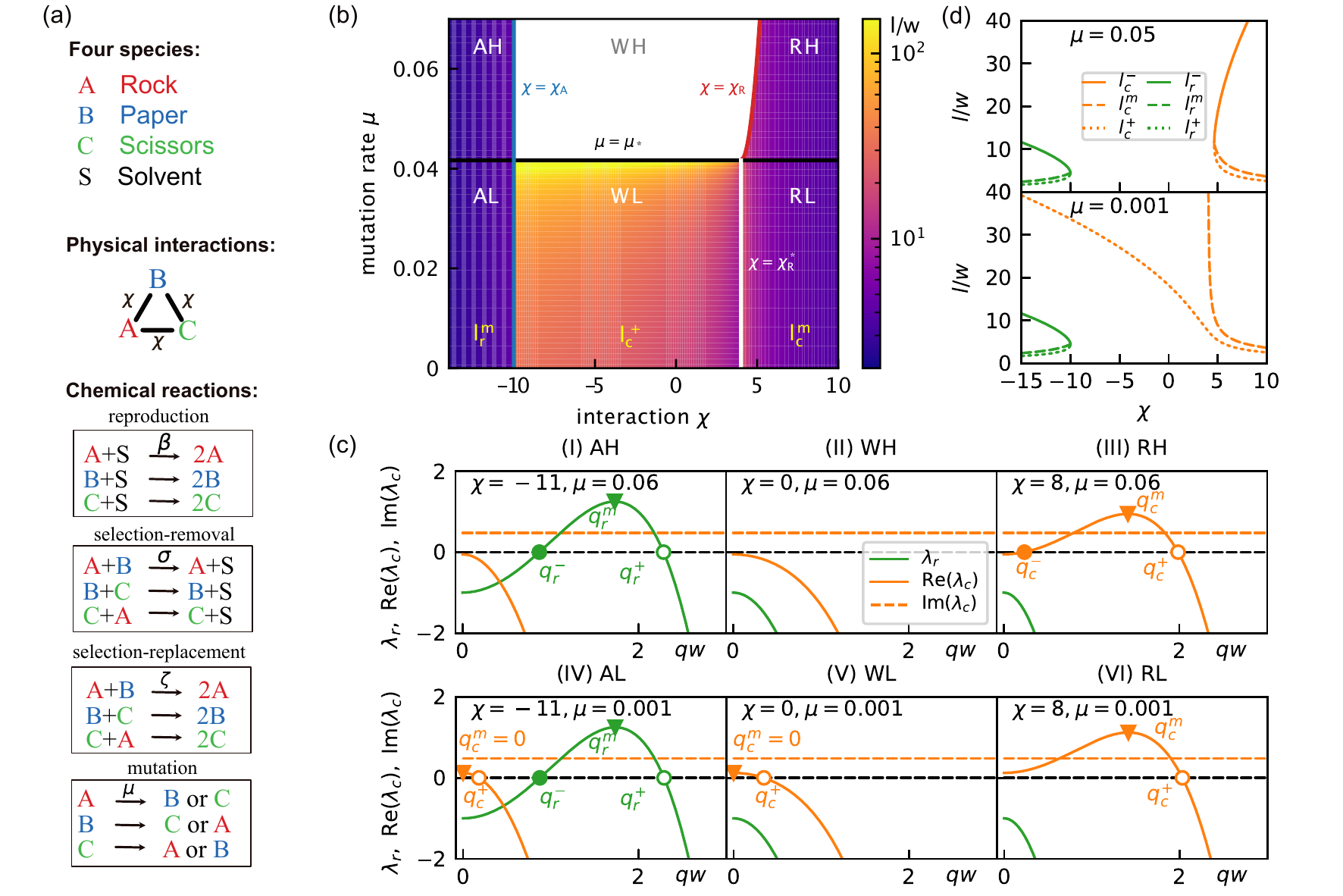}
  \caption{ 
  \textbf{Linear stability analysis reveals distinct parameter regions.}
  (a) Schematic of physical interactions and chemical reactions of three species $A$, $B$, $C$, and the inert solvent $S$.
  (b) Stability diagram distinguishing regions of low (L) and high (H) mutation rate~$\mu$ as well as strong attraction (A), weak interaction (W), and strong repulsion (R).
  The critical lines follow from \Eqref{eq:mustar} (black line), \Eqref{eq:chiA} (blue line), \Eqref{eq:chiR} (red line), and $\chiRideal = 1/\phi_0$ (white line).
  The colors represent the length scales $\lrm$ in regimes AH and AL, $\lcm$ in RH and RL, and $\lc^+$ in WL. 
  (c) Representative dispersion relations $\lambda(q)$ in the six regimes.
  Green curves represent real eigenvalues $\EVr(q)$ with left root ($\qr^-$, green disk), right root ($\qr^+$, green circle), and maximum ($\qrm$, green triangle) marked.
  Solid orange curves and symbols represent the respective values for the real part of the complex eigenvalues, whereas the dashed orange lines mark the imaginary part  $\Im(\EVc)=\omega_*$; see \Eqref{eq:omegaH}. 
  (d) Typical length scales as a function of $\chi$ at $\mu=0.05>\mu_*$ (upper panel) and $\mu=0.001<\mu_*$ (lower panel).
  The subscript and superscript of the length scale $l$ correspond to those of wavenumber $q$ in (c) using $l=2\pi/q$.
   (b--d) Additional model parameters are $\beta=\sigma=D=1$ and $\zeta=0.6$.
   }
  \label{fig:overall} 
\end{figure*}

We consider an incompressible, isothermal fluid comprising three species $A$ (rock), $B$ (paper), and $C$ (scissors) as well as an inert solvent $S$.
This system is described by the volume fractions $\phi_A(\vect r,t)$, $\phi_B(\vect r,t)$, and $\phi_C(\vect r,t)$, where $\vect r$ is the spatial position and $t$ is time, and the solvent occupies the remaining fraction $\phi_S=1-(\phi_A+\phi_B+\phi_C)$.
We explicitly include physical interactions and chemical reactions among the species in our model; see \figref{fig:overall}(a).

\subsubsection{Physical interactions}
We describe physical interactions using thermodynamics based on the Flory-Huggins free energy \cite{safran2018statistical, rubinstein2003polymer, Cahn1958},
\begin{multline}
  \label{eq:free_energy}
    F\left[\phi_A,\phi_B,\phi_C\right] =
    \frac{k_\mathrm{B}T}{\nu}\int\Bigl[
    \chi(\phi_A\phi_B+\phi_A\phi_C+\phi_B\phi_C)
\\+
    \hspace{-1em}\sum_{i=A,B,C,S}\hspace{-1em}\phi_i\ln\phi_i
+\frac{1}{2}w^2\hspace{-0.8em}\sum_{j=A,B,C}\hspace{-0.8em}\bigl|\nabla\phi_j\bigr|^2
\Bigr]\diff \vect r
\;,
\end{multline}
where the integral is over the volume of the system, $k_\mathrm{B} T$ is the relevant energy scale, and $\nu$ denotes the molecular volume, which is the same for all species for simplicity.
The first term in the square bracket describes the physical interactions among the species $A$, $B$ and $C$, the second term captures translational entropies of all four species, and the last term limits the width of interfaces between coexisting phases to roughly $w$ in strongly interacting systems~\cite{Cahn1958}.
The physical interactions are quantified by the Flory parameter~$\chi$:
Positive~$\chi$ denotes repulsion, whereas negative~$\chi$ represents attraction.
For simplicity, we only consider symmetric interactions, i.e., the same value of $\chi$ for all pairs of $A$, $B$ and $C$, while the solvent is inert, but in general the value could be species-dependent.

The free energy defined in \Eqref{eq:free_energy} allows for inhomogeneous equilibrium states when the physical interactions are sufficiently strong~\cite{Cahn1958, Mao2018, Zwicker2022a}.
In particular, a phase enriched in species $A$, $B$, and $C$ segregates from the solvent $S$ for strong attraction ($\chi < \chiAideal$), whereas strong repulsion ($\chi > \chiRideal$) leads to three phases which are each enriched in one of the species and the solvent.
In the case where the species $A$, $B$, and $C$ have equal average fraction $\phi_0$, the critical values are 
\begin{align}
	\label{eqn:ciritical_chi}
	\chiAideal &=-\frac{1}{2\phi_0(1-3\phi_0)}
& \text{and} &&
	\chiRideal &= \frac{1}{\phi_0}
	\;,
\end{align}
which follows from a linear stability analysis shown in the Appendix.
Taken together, we expect that the two critical values given in \Eqref{eqn:ciritical_chi} separate three qualitatively different regions in parameter space.

\subsubsection{Cyclic dominant chemical reactions}
Following previous rock-paper-scissors game studies~\cite{szczesny2013does,szczesny2014characterization}, we consider general chemical reactions that include reproduction, selection, and mutation; see \figref{fig:overall}(a).
Reproduction happens with rate $\beta$ when a species $i \in \{A, B, C\}$ meets solvent, which could also play a role similar to empty space.
Selection comes in two variants, which both encode the typical rock-paper-scissors rules, where species~$i$ dominates species~$i+1$ while being dominated by species~$i-1$, using the cyclically ordered index such that $A+1=B$, $B+1=C$, and $C+1=A$.
The first selection variant removes the dominated species with rate~$\sigma$, whereas the second variant replaces the dominated species by the dominating one with rate~$\zeta$ in a zero-sum process.
Finally, random mutations happen with rate $\mu$.
Combining all these processes, the reaction rate of species $i$ reads
\begin{align}
  \label{eq:chemical_reaction}
R_i&=\phi_i\bigl[\beta\phi_S-\sigma\phi_{i-1}+\zeta(\phi_{i+1}-\phi_{i-1})\bigr]
\notag \\
&\quad+\mu(\phi_{i-1}+\phi_{i+1}-2\phi_i)
\end{align} 
for $i \in \{A, B, C\}$ with positive rates $\beta$, $\sigma$, $\zeta$, and $\mu$. %
For  $\beta=\sigma=\mu=0$, the model reduces to the cyclic Lotka-Volterra model with equal replacement rate $\zeta$ \cite{nash1950equilibrium,reichenbach2006coexistence}, whereas $\zeta=\mu=0$ leads to the May-Leonard model \cite{may1975nonlinear}.

In the simplest case without spatial dependence the dynamics of the three species are given by $\partial_t \phi_i = R_i$.
This system undergoes a supercritical Hopf bifurcation when $\mu$ decreases below $\mu_*$, where \cite{szczesny2014coevolutionary}
\begin{eqnarray}
  \label{eq:mustar}
  \mu_*=\frac{\beta\sigma}{6(3\beta+\sigma)}
  \;,
\end{eqnarray} and develops a stable limit cycle with frequency 
\begin{eqnarray}
  \label{eq:omegaH}
  \omega_*=\frac{\sqrt{3} \beta  (2 \zeta +\sigma )}{2 (3 \beta +\sigma )}
  \;.
\end{eqnarray}
Combining these oscillating reactions with ideal diffusion then leads to steady spiral waves and other oscillating states~\cite{nagatani2020diffusively,szczesny2013does,szczesny2014characterization,mobilia2010oscillatory}.
However, it is unclear how physical interactions affect the oscillating states and how the chemical reactions modify the equilibrium behavior of phase separation.

\subsubsection{Combined model}
To combine physical interactions and chemical reactions, we use the exchange chemical potentials
\begin{eqnarray}
   \bar \mu_i=\frac{\nu}{k_\mathrm{B}T} \, \frac{\delta F}{\delta \phi_i}
    \;,
\end{eqnarray}
to express diffusive fluxes~$\vect j_i= -\Lambda_i \nabla \bar\mu_i$ in the continuity equation $\partial_t \phi_i = \nabla.\vect{j}_i + R_i$~\cite{julicher2018hydrodynamic,de1984non}.
Hence,
\begin{eqnarray}
    \label{eq:dynamics}
	\partial_t \phi_i=\nabla\cdot[D_i\phi_i\nabla \bar\mu_i]+R_i
	\;,
\end{eqnarray}
where $R_i$ is given by \Eqref{eq:chemical_reaction}.
Here, $D_i$ are the diffusivities of the species $i=A, B, C$, which are related to mobilities $\Lambda_i=D_i\phi_i$ in this multicomponent system~\cite{Kramer1984}. %

To analyze the behavior of  \Eqref{eq:dynamics}, we first use linear stability analysis to identify qualitatively different regimes and associated length scales of patterns.
We then study the dynamical behavior in detail using numerical simulations in a two-dimensional system with periodic boundary conditions.
In the simulations, we choose $\beta=1$ to set the time scale, and $w=1$ to set the length scale. 
For simplicity, we also set $D_i=D=1$ for all three species to focus on how the physical interaction parameter $\chi$, the mutation rate~$\mu$, and replacement rate~$\zeta$ affect the pattern formation and cyclic behavior of the system.

\subsection{Linear stability analysis reveals phase diagram}
\label{sec:lsa}
To reveal the basic behavior of the model, we first analyze the stability of the only uniform steady state of \Eqref{eq:dynamics}, which is $\phi_i(\vect r)=\phi_0$ with
\begin{eqnarray}
  \label{eq:uniform_solution}
	\phi_0=\frac{\beta}{3\beta+\sigma}
	\;.%
\end{eqnarray}
We focus on the case of an equal average fraction~$\phi_0=\frac14$ for $A$, $B$, $C$, and $S$, implying $\sigma=\beta$.
In the linear regime of small perturbations, we assess the stability of this homogeneous state by evaluating the growth rates~$\lambda$ of harmonic perturbations with wave number~$q$; see Appendix.
For each $q$, we obtain three eigenvalues of the Jacobian matrix associated with \Eqref{eq:dynamics}, of which one is always real (denoted by $\EVr$), whereas the remaining two eigenvalues are complex conjugates of each other, denoted as $\Re(\EVc) \pm \Im(\EVc)$.
The homogeneous state is unstable if any eigenvalue has a positive real part and the associated imaginary part represents the oscillation frequency, which is related to $\omega_*$ given by \Eqref{eq:omegaH}.
Note that $\EVr$ is independent of the mutation rate~$\mu$, whereas the stability of the complex modes depends on $\mu$.
In particular, they are stable in the limit of long wavelengths, $\mathrm{Re}(\EVc(q=0))<0$, if and only if the mutation rate $\mu$ is higher than the critical value $\mu_*$ given by \Eqref{eq:mustar}, which clearly distinguishes a regime of low and high mutation rate, which we denote by L and H, respectively.
Within each region, we can furthermore distinguish regions of strong attraction (region A), weak interaction (region W), and strong repulsion (region R), based on the critical values suggested by \Eqref{eqn:ciritical_chi}.
The combination of these two characteristics leads to the six distinct parameter regimes shown in \figref{fig:overall}, which we will now discuss in more detail.

In region WH with \emph{weak} interactions ($\chiA<\chi<\chiR$) and \emph{high} mutation rates ($\mu>\mu_*$), the uniform solution \Eqref{eq:uniform_solution} is stable, since the real parts of all eigenvalues are negative; see \figref{fig:overall}(c)(II).
The critical values for the physical interactions,
\begin{subequations}
 \begin{align}
  \label{eq:chiA}
  \chiA &=\chiAideal-w\sqrt{\frac{3\beta+\sigma}{D}}
	\qquad\qquad \text{and}\\
  \label{eq:chiR}
  \chiR(\mu) &=\chiRideal+w\sqrt{\frac{2\sigma (\mu/\mu_*-1)}{D}}
  \;,
 \end{align}
\end{subequations}
follow from solving $\mathrm{max}(\EVr)=0$ and $\mathrm{max}(\mathrm{Re}(\EVc))=0$ for $\chi$, respectively.
Here, we used \Eqref{eq:uniform_solution} to compare to $\chiAideal$ and $\chiRideal$ given by \Eqref{eqn:ciritical_chi}, which mark the influence of phase separation.
Consequently, chemical reactions shift both critical values to stronger interactions, consistent with reactions suppressing phase separation~\cite{ziethen2022nucleation}.
 
In region AH with strong \emph{attraction} ($\chi<\chiA$) and \emph{high} mutation rate ($\mu>\mu_*$), the real eigenvalue $\EVr$ is positive if $\qr^-<q<\qr^+$, where $\qr^{-}$ ($\qr^+$) is the left (right) root of $\EVr$; see \figref{fig:overall}(c)(I).
The wavelength~$\lrm=2 \pi /\qrm$ of the corresponding instability can be estimated from the wave number~$\qrm$ of the most unstable mode and reads
\begin{equation}
    \label{eq:lrm}
    \lrm =2 \pi w\left[-\frac{2 \beta  \sigma }{2 \beta  \sigma  \chi +(3 \beta +\sigma )^2}\right]^{\frac12}
    \;.
\end{equation}
Consequently, $\lrm$ decreases slightly for smaller physical interaction $\chi$; see the dashed green curve in the upper panel of \figref{fig:overall}(d).
We thus expect stationary pattens with length scales close to $\lrm$ in region AH.

In region RH with strong \emph{repulsion} ($\chi>\chiR$) and \emph{high} mutation ($\mu>\mu_*$), the complex eigenvalues $\EVc$ exhibit an instability for $\qc^-<q<\qc^+$; see \figref{fig:overall}(c)(III).
The associated  most unstable wavelength~$\lcm=2 \pi/\qcm$ reads
\begin{equation}
    \label{eq:lcm}
    \lcm=2\pi w\left[\frac{2 }{\chi-(3 +\sigma /\beta)}\right]^{\frac12}
    \;,
\end{equation}
and decreases for stronger repulsion; see the dashed orange curve in the upper panel of \figref{fig:overall}(d).
Since the imaginary parts for these modes are nonzero, we expect oscillating patterns with length scales close to $\lcm$.

In region AL with strong \emph{attraction} ($\chi < \chiA$) and \emph{low} mutation rate ($\mu<\mu_*$), we find the same unstable real modes as in region AH as well as additional unstable complex modes for $0<q<\qc^+$, although their maximal growth rate is typically smaller than that of the real modes.
However, linear stability analysis does not provide any information on how these modes interact and we thus expect a rich behavior in this region.

 In region WL with \emph{weak} interaction ($\chiA< \chi<\chiR$) and \emph{low} mutation rate ($\mu<\mu_*$), the oscillating modes are unstable for $q<\qc^+$, whereas $\EVr<0$; see \figref{fig:overall}(c)(V).
 The length scale of the most unstable mode diverges ($\qcm=0$), so the length scale $\lc^+=2\pi /\qc^+$ associated with the largest unstable wave number~$\qc^+$,
\begin{equation}
  \label{eq:lcplus}
  \lc^+=2 \pi w \left(\frac{2 }{  \tilde{\chi} + \Bigl[\tilde{\chi} ^2 + \frac{2  \sigma  w^2}{D}\bigl(1-\frac{\mu}{\mu_*}\bigr)\Bigr]^{1/2}}\right)^{\frac12}
  \;,
 \end{equation}
with $\tilde{\chi}=\chi-3-\sigma/\beta$, is most relevant.
This length scale decreases significantly as $\chi$ increases; see dotted orange line in lower panel of \figref{fig:overall}(d).
  
Finally, in region RL with strong \emph{repulsion} ($\chi > \chiR$) and \emph{low} mutation rate ($\mu<\mu_*$), we find the same unstable modes as in region WL, but the length scale~$\lcm=2\pi/\qcm$ of the most unstable mode is now finite.
This length scale decreases for larger interaction parameters $\chi$; see dashed orange curve in lower panel of \figref{fig:overall}(d).
We distinguish the regions WL and RL based on whether $\qcm$ is zero or not, which provides the critical physical interaction $\chiR^*=\phi_0^{-1}$.
The fact that this threshold value is identical to $\chiRideal$ given by \Eqref{eqn:ciritical_chi} suggests that the transition is governed by phase separation induced by the physical interactions.

\begin{figure*}
  \centering
  \includegraphics[width=\textwidth]{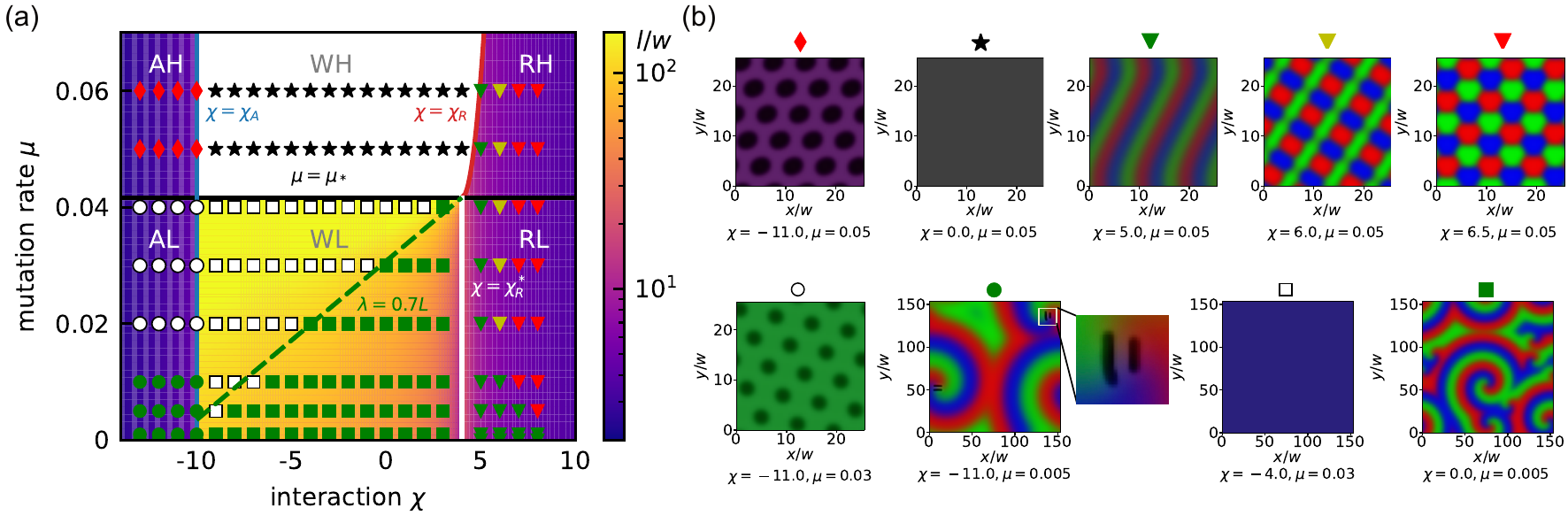}
  \caption{\textbf{Numerical simulations reveal diverse patterns.}
  (a) Phase diagram with stability lines copied from \figref{fig:overall}(b).
  Background colors also correspond to \figref{fig:overall}(b), except in region WL, where they mark the length scale~$\lambda$ given by \Eqref{eq:lambda}. %
  The green dashed lines corresponds to $\lambda \approx 0.7\,L$, which fits the transition best.
  Symbols classify different patterns corresponding to examples in panel (b).
  (b) Snapshots of representative patterns for panel (a).
  Colors represent the abundance of the three species using RGB triplets: (red, green, blue) = ($\phi_A$, $\phi_B$, $\phi_C$).
 We used $L=25.6 \, w$, $\Delta x=0.4\, w$, and $t=10^5 \beta$ for the snapshots %
 in the first row and the one marked with white circle in the second row.
 Movies of these states are enclosed with the Appendix.
 (a--b) Model parameters are $D=\beta=\alpha=1$ and $\zeta=0.6$.
  Simulation parameters are $L/w=153.6$ with discretization $\Delta x/w=0.6$ and we evaluate patterns after time $t=10^4\beta$.
 } 
  \label{fig:snapshot} 
\end{figure*}

Taken together, linear stability analysis provides a qualitative picture of the five unstable regimes, and it predicts the associated critical curves; see \figref{fig:overall}(b).
The analysis also provides typical length scales in different regimes; see color shading in \figref{fig:overall}(b) and \figref{fig:overall}(d).
We next corroborate the phase diagram with detailed simulations and analyze the non-linear behavior of the model.
For simplicity, we consider two-dimensional simulations in square boxes of side length~$L$ with periodic boundary conditions and we implement the spatial derivatives using finite differences~\cite{Zwicker2020}.
The simulation results summarized in \figref{fig:snapshot} indicate that the uniform state is indeed stable in region WH (black stars), whereas complex patterns emerge in the unstable regimes, which we discuss in detail in the following sections.

\subsection{Weak interactions affect length scales, but not frequency, of spiral waves}
We start by discussing weak physical interactions, where we expect qualitatively similar behaviors to systems without interactions.
In the region WL with low mutation rates, where patterns actually form, we observe two main types of oscillating patterns:
Homogeneous oscillations (white squares in upper left part of region WL in \figref{fig:snapshot}) or spiral waves (green squares in lower right part of the region WL), which are expected from the linear stability analysis.
This raises the question of why spiral waves are apparently suppressed for parameters above the diagonal green dashed line in \figref{fig:snapshot}.

\begin{figure*}
    \centering
    \includegraphics[width=\textwidth]{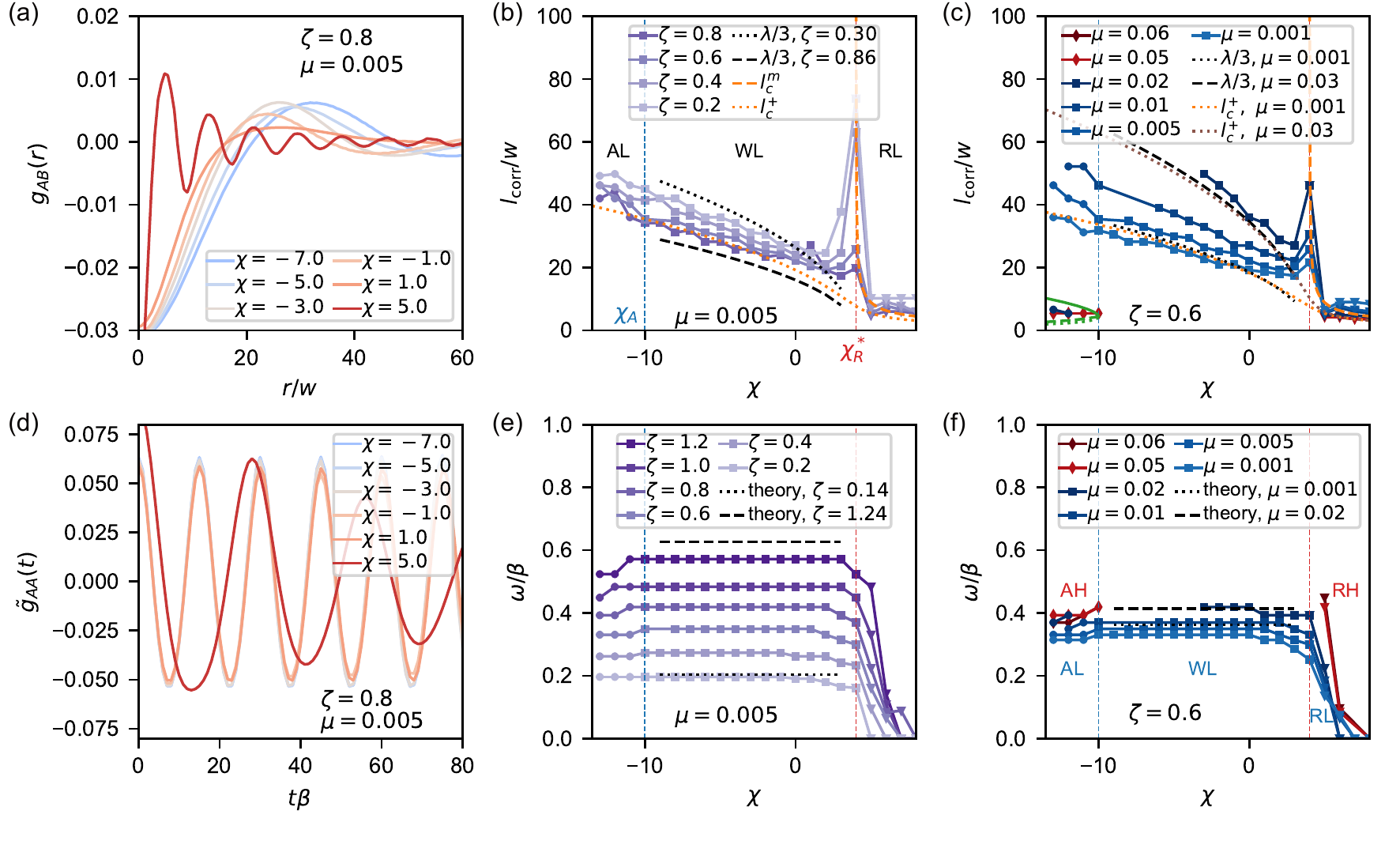}
    \caption{\textbf{Length- and time-scales of dynamic patterns.}
	(a) Spatial correlation function $g_{AB}(r)$ as a function of distance $r$ for various physical interactions~$\chi$ at $\zeta=0.8$ and $\mu=0.005$.
	(b) Correlation length scale $l_\mathrm{corr}$ determined from first maximum of $g_{AB}(r)$ as a function of $\chi$ at $\mu=0.005$.
	The black dashed (dotted) line corresponds to $\frac13\lambda$ calculated from \Eqref{eq:lambda} at $\zeta=0.86$ ($\zeta=0.30$).
	The orange lines are the same as in \figref{fig:overall}(d) at $\mu=0.005$.
	(c) $l_\mathrm{corr}$ as function of $\chi$ for $\zeta=0.6$.
	The black dashed (dotted) line corresponds to $\frac13\lambda$ calculated by \Eqref{eq:lambda} at $\mu=0.03$ ($\mu=0.001$).
	The green and orange lines are the same as in \figref{fig:overall}(d) at $\mu=0.001$.
	The brown dotted line is $\lc^+$ at $\mu=0.03$.
	(d) Temporal correlation function $\tilde{g}_{AA}(t)$ as a function of lag time $t$ for various $\chi$ at $\zeta=0.8$ and $\mu=0.005$.
	(e) Frequency $\omega$ determined from the first maximum of $\tilde{g}_{AA}$ as a function of $\chi$ for $\mu=0.005$.
	The black dashed (dotted) line shows $\omega$ given by \Eqref{eq:omega} for $\zeta=1.24$ ($\zeta=0.14$).
	(f) $\omega$ as a function of $\chi$ for $\zeta=0.6$.
	The black dashed (dotted) line shows $\omega$ given by \Eqref{eq:omega} at $\mu=0.02$ ($\mu=0.001$).
	(a--f) The vertical dashed blue (red) lines denote the critical interactions $\chiA=-10$ ($\chiR^*=4$). %
	Additional model parameters are $\beta=\alpha=D=1$.}
    \label{fig:lengthscale_frequency} 
  \end{figure*}

To address this question, we first carefully analyze the regime with spiral waves.
We quantify the wavelength of the spiral waves using the static spatial correlation function $g_{\alpha\beta}(r)\equiv g_{\alpha\beta}(|\br-\br'|)=\langle \phi_{\alpha}(\br) \phi_{\beta}(\br')\rangle-\langle\phi_{\alpha}(\br)\rangle\langle\phi_{\beta}(\br')\rangle$ from simulated snapshots.
\figref{fig:lengthscale_frequency}(a) shows the cross-correlation between $A$ and $B$, allowing us to define the correlation length $\lCorr$ as the position of the first peak of $g_{AB}$.
\figref{fig:lengthscale_frequency}(b--c) show that $\lCorr$ generally decreases with increasing interaction parameter~$\chi$ for $\chiA<\chi<\chiR^*$, implying that stronger repulsion between species shortens the length scales of spiral waves.

To understand the effect of physical interaction on spiral waves, we next use a multiscale expansion around the Hopf bifurcation $\mu=\mu_*$, to map \Eqref{eq:dynamics} to a complex Ginzburg-Landau equation (CGLE) with real diffusion coefficient \cite{szczesny2013does}; see Appendix.
The CGLE also exhibits spiral waves, so we can use established theory~\cite{aranson1993theory,aranson2002world} to predict their wavelength $\lambda$,
\begin{equation}
    \label{eq:lambda}
    \lambda=2 \pi \sqrt{\frac{D(1-\frac{\chi}{3+\sigma/\beta})}{3(\mu_*-\mu)(1-|U|^2)}}
    \;,
\end{equation}
where $|U|^2$ is the square of the amplitude of the solution of the CGLE, which only depends on $\zeta$ for fixed $\beta$ and $\sigma$; see Appendix.
\Eqref{eq:lambda} shows that the wavelength~$\lambda$ decreases for larger $\chi$, and \figref{fig:lengthscale_frequency}(b--c) show that the expression is close to our numerical estimates, even though $\mu$ is not very close to $\mu_*$.
\Eqref{eq:lambda} also predicts that smaller mutation rates $\mu$ lead to shorter wavelengths, consistent with \figref{fig:lengthscale_frequency}(c) and a previous study \cite{szczesny2013does}. %
Interestingly, the length scale $\lc^+$ given in \Eqref{eq:lcplus} %
also describes the observed behavior accurately; see \figref{fig:lengthscale_frequency}(b--c).
In fact, we find $\lim_{\mu\rightarrow \mu_*}\lc^+ = \sqrt{1-|U|^2}\lambda$ close to the Hopf bifurcation.
Finally, increasing the replacement rate~$\zeta$ leads to smaller amplitudes $|U|$ and thus decreased wavelengths; see \figref{fig:lengthscale_frequency}(b) and Appendix.
Note that we also observe patterns that are reminiscent of the Eckhaus and absolute instability of the CGLE~\cite{szczesny2013does,szczesny2014characterization} at large replacement rates~$\zeta$; see \figref{fig:S_snapshots_zeta18}. 
Taken together, we found that the mapping to the CGLE provides a faithful theoretical prediction of the length scales of spiral waves as a function of the relevant model parameters. 

The dependence of the length scale~$\lambda$ of the spiral waves prompted us to hypothesize that spiral waves can only emerge when their intrinsic length scale is smaller than the system size.
Indeed, the green dashed line in \figref{fig:snapshot} indicates that spiral waves only emerge when $\lambda \lesssim L$.
We thus conclude that the cases where we observe homogeneous oscillations would show spiral waves in larger systems.

We next quantify the frequency~$\omega$ of the oscillating patterns using the first peak of the temporal correlation function $\tilde{g}_{AA}(t)=\tilde{g}_{AA}(|t_1-t_2|)=\langle \phi_A(\br,t_1) \phi_A(\br,t_2)\rangle-\langle\phi_A(\br,t_1)\rangle\langle\phi_A(\br,t_2)\rangle$; see \figref{fig:lengthscale_frequency}(d).
\figref{fig:lengthscale_frequency} shows that the interaction parameter~$\chi$ hardly affects $\omega$ in the weak interaction regime ($\chiA<\chi<\chiR^*$).
We rationalize this behavior by mapping \Eqref{eq:dynamics} to a reaction-diffusion equation in the limit of weak interactions~$\chi$, revealing that $\chi$ only affects cross-diffusion, but not the reactions; see Appendix.
The associated frequency~$\omega_*$ of the most unstable mode is given by \Eqref{eq:omegaH} and explains most of the behavior of the numerically determined~$\omega$.
However, $\omega_*$ does not depend on the mutation rate $\mu$, so this approximation cannot explain the dependence of $\omega$ on $\mu$.
To capture this phenomenologically, we use the mapping to the CGLE presented in the Appendix, which provides a correction,
\begin{equation}
  \label{eq:omega}
  \omega=\omega_* - 3(\mu_*-\mu)c|U|^2
  \;,
\end{equation} 
where $c$ is a constant depending on $\beta$, $\sigma$, and $\zeta$; see \Eqref{eq:c_zeta} in the Appendix.
This expression correctly predicts that $\omega$ is independent of $\chi$ and that it increases for larger~$\zeta$ and~$\mu$; see \figref{fig:lengthscale_frequency}(e--f).

Taken together, we find that weak repulsion in region WL shortens the wave length of spiral waves, while their period is unaffected.
A multiscale expansion around the Hopf bifurcation leads to a CGLE, which reveals that this behavior is caused by cross-diffusion resulting from physical interactions, analogously to the effect of weak interactions on Turing patterns~\cite{menou2023physical}.

\subsection{Oscillations and phase separation coexist for strong attractive interactions}

We next focus on systems with strong attraction ($\chi< \chiA$), where we first consider weak mutation rates ($\mu<\mu_*$, region AL).
We expect that the spiral waves we found for weak attractions persist, albeit with longer wave lengths, following the observed trend in region WL.
Indeed, \figref{fig:snapshot} demonstrates spiral waves at low mutation rate (green circle), and \figref{fig:lengthscale_frequency} confirm that the length scale increases for smaller~$\chi$ while the frequency stays almost constant.
Moreover, the effects of the mutation rate~$\mu$ and the replacement rate~$\zeta$ are similar in regions WL and AL.
However, we also observe that spiral waves form in a larger parameter region than expected: In region WL, boundary effects suppressed spiral waves that are comparable to or larger than the system size (white symbols above the green dashed line in \figref{fig:snapshot}), while this suppression is apparently much weaker in region AL.
Since this transition coincides with the line $\chi=\chiA$, we hypothesize that strong attractive interactions stabilize spiral waves.

Strong attraction can lead to phase separation, where the three species $A$, $B$, and $C$ co-segregate from the solvent~$S$.
Indeed, the dark spots in the snapshots shown in \figref{fig:snapshot}(b) correspond to solvent-rich droplets, which are absent in region WL.
Interestingly, these solvent droplets co-localize with defect cores of spiral waves.
On the one hand, this suggests that phase separation can only proceed in the relatively calm defect cores while the comparatively strong spiral waves prevent phase separation by mixing the system effectively.
Indeed, spatiotemporal chaos at large replacement rates~$\zeta$ can prevent the formation of solvent droplets close to the transition ($\chi \lesssim \chiA$); see Fig. 4 in the Appendix.
On the other hand, the solvent droplets formed by phase separation apparently stabilize spiral waves, similar to rigid obstacles~\cite{eckstein2020spatial,zykov2018spiral,davidenko1992stationary,pertsov1993spiral}.
Taken together, positive feedback between formation of solvent droplets and spiral waves apparently stabilizes this state even if the system would otherwise be too small.

For larger mutation rates~$\mu$, spiral waves are absent even if $\mu<\mu_*$.
Presumably, this is again caused by limitations imposed by the system size, consistent with the increasing pattern length scale shown in \figref{fig:lengthscale_frequency}(c).
When spiral waves are absent, phase separation can take place everywhere and we observe a regular hexagonal lattice of solvent droplets embedded in a phase enriched in the other species; see snapshot labeled by a white disk in \figref{fig:snapshot}.
For stronger attraction, we also sometimes observe bicontinuous structures with a fixed length scale; see Fig.~2 in the Appendix.
In both cases, coarsening is suppressed by reactions~\cite{Zwicker2015}, and the correlation length scale $l_\mathrm{corr}$ is within the band of unstable real modes ($\qr^-<q<\qr^+$) predicted by the linear stability analysis; see the red symbols and green curves in \figref{fig:lengthscale_frequency}(c).
Moreover, \figref{fig:lengthscale_frequency}(f) shows that the large connected phase oscillates between the three species $A$, $B$, and $C$ with a frequency close to $\omega_*$, consistent with the prediction of the frequency of the complex mode.
Taken together, linear stability analysis predicts the most important properties of the hexagonally arranged solvent droplets embedded in an oscillating phase in region AL. 

\begin{figure*}
  \centering
  \includegraphics[width=\textwidth]{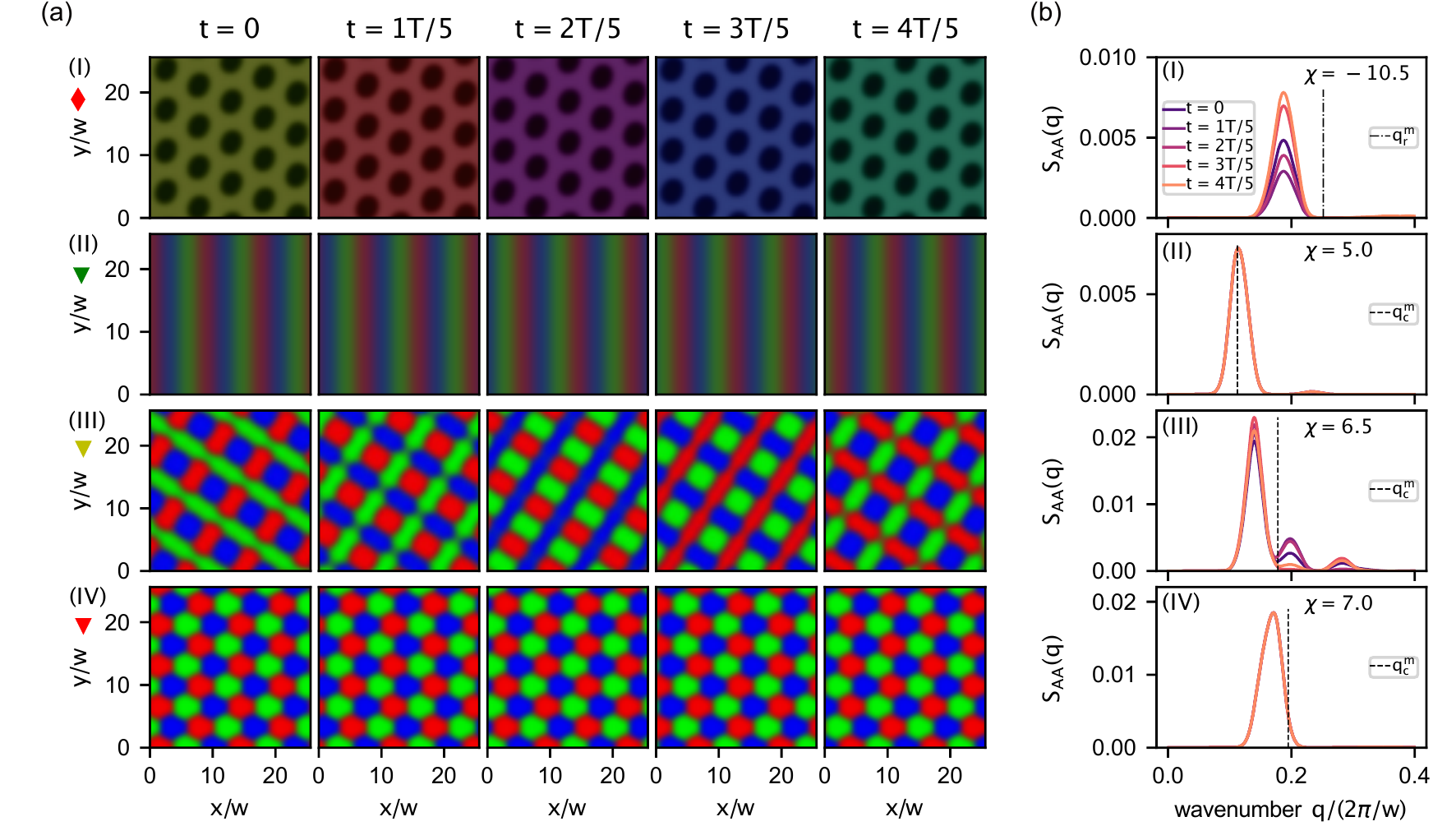}
  \caption{\textbf{Details of complex, oscillating patterns.}
  (a) Snapshots showing one temporal period for four physical interactions ($\chi=-10.5, 5, 6.5, 7$; top to bottom) at $\mu=0.05>\mu_*$ and $\zeta=1$.
  The corresponding periods are $T\approx 11\beta, \ 11\beta,\ 52\beta,\ 610\beta$.
  (b) Structure factors $S_{AA}(q)$ corresponding to data in (a).
  The vertical lines mark the wave number of the most unstable mode determined from linear stability analysis.
  }
  \label{fig:structure_factor}
\end{figure*}

Linear stability analysis predicts that oscillations cease once the the mutation rate~$\mu$ becomes larger than $\mu_*$.
However, our numerical simulations of the full model show that the states do not change qualitatively when we cross this stability boundary:
The hexagonal pattern of solvent droplets remains and the connected phase still oscillates between the three species; see \figref{fig:structure_factor}(a)(I).
While this behavior is obviously driven by non-linear effects, the length scale of the hexagonal pattern still decreases for decreasing~$\chi$ and increasing~$\mu$, consistent with the trend predicted by linear stability analysis.

We conclude that the competition of the Turing instability and the Hopf instability governs the behavior for strong attraction ($\chi<\chiA$).
For low $\mu$ and sufficiently large systems, we observe spiral waves with solvent droplets at their core, whereas hexagonal patterns of solvent droplets embedded in an oscillating phase emerge for larger $\mu$ and in small systems.
Both behaviors are impossible in excitable systems with ideal diffusion, demonstrating the qualitatively new effects that strong attraction between species can bring.

\subsection{Strong repulsion leads to oscillating lattices}

Finally, we discuss strong repulsion between species ($\chi>\chiR$), where we predict a segregation of the species $A$, $B$, and $C$ from each other while the solvent is homogeneously distributed.
The linear stability analysis shown in \figref{fig:overall} predicts that complex modes are unstable for all values of the mutation rate $\mu$, whereas the critical value~$\mu_*$ merely governs the stability of homogeneous perturbations ($q=0$).
Consequently, we expect oscillatory patterns in both the regions RH and RL.

Our numerical simulations shown in \figref{fig:structure_factor}(a) reveal oscillating patterns for strong repulsion.
For interaction strengths~$\chi$ close to the critical value~$\chiR$, the corresponding frequency~$\omega$ is comparable to the value $\omega_*$ predicted by \Eqref{eq:omegaH}, but $\omega$ drops strongly with increasing repulsion~$\chi$; see \figref{fig:lengthscale_frequency}(e,~f).
Concomitantly, the spatial patterns change: 
Close to the transition, we find oscillating stripes; see snapshots marked by green triangles in \figref{fig:snapshot} and \figref{fig:structure_factor}(a).
As $\chi$ increases, the stripes first transition to slowly oscillating square lattices (marked by yellow triangles) and then further to slowly oscillating hexagonal lattices (marked by red triangles).
The length scales~$l_\mathrm{corr}$ of these patterns are comparable to the length scales $\lcm$ of the most unstable mode, which also captures the observation that larger repulsion~$\chi$ leads to smaller structures; see \figref{fig:lengthscale_frequency}(a--c).
However, the observed increase of $l_\mathrm{corr}$ with decreasing $\mu$ and decreasing~$\zeta$ cannot be explained by $\lcm$ and thus likely results from non-linear effects.
Moreover, for square and hexagonal lattices, $l_\mathrm{corr}$ is a bit larger than predicted from linear stability analysis, consistent with results in reaction-diffusion systems~\cite{yang2004stable}. 
Finally, the spatiotemporal chaos emerging at large replacement rates~$\zeta$ can prevent the formation of regular patterns close to the transition ($\chi \gtrsim \chiR$); see Fig. 4 in the Appendix.
Taken together, this rich behavior indicates that strong repulsive interactions affect pattern formation strongly, presumably because repulsion segregates the species from each other so that the cyclic-dominant reactions are most active at interfaces.

\section{Discussion}

We investigated the behavior of three species that interact physically and exhibit cyclic dominant reactions to study the effect of physical interaction on spatiotemporal patterns.
For weak interactions, the mapping to the complex Ginzburg-Landau equation (CGLE) reveals that interactions mainly cause cross-diffusion, which affects length-scales, but not time-scales, of the resulting spiral waves.
In contrast, qualitatively new patterns emerge if interactions are strong:
Strong attraction leads to phase separation of the solvent from all species, which exhibit spiral waves or oscillations.
In this case, the typical coarsening of passive phase separation is suppressed, droplets can stabilize spiral waves, and oscillations appear even without a complex unstable mode in the linear stability analysis.
Conversely, for strong repulsion, all species segregate from each other, limiting chemical interactions to interfaces, which results in various oscillating lattices.
In summary, we find that linear stability analysis and the mapping to the CGLE explain the influence of weak interactions, whereas these approaches are less predictive for the qualitatively different patterns emerging for strong interactions.

Cyclic dominant reactions have been linked to biodiversity in ecological contexts~\cite{reichenbach2007mobility, tilman1994competition}, where the interplay of species and their respective survival impacts biodiversity.
Our analysis suggests that repulsive interactions between species result in spatiotemporal patterns even for large mutation rates~$\mu$, where otherwise a single species would dominate.
Conversely, attraction between species favors co-localization and the resulting competition makes extinction more likely.
Moreover, physical interactions impact resulting dynamics qualitatively, suggesting that ecological patterns are affected and interactions need to be included when studying biodiversity.

To build a general understanding of the impact of interactions in realistic systems, we will need to consider more complex models.
For instance, we could consider more complex chemical reactions, e.g., including death rates~\cite{islam2022effect} or non-symmetric reactions~\cite{berr2009zero}, although some of the complexity might simply induce a renormalization of parameters \cite{peltomaki2008three}.
In contrast, more diverse physical interactions can provide additional states already in equilibrium phase separation~\cite{Zwicker2022,Mao2018}.
In particular, considering more than three species provides room for additional patterns~\cite{peltomaki2008three,bayliss2020beyond,hu2022emergent}, and we suspect that the lattices we observed at strong repulsion will look completely different.
Realistic systems will also exhibit stochasticity~\cite{szolnoki2014cyclic} and spatial heterogeneity~\cite{szolnoki2020pattern}, which sometimes can be approximated by considering networks~\cite{postlethwaite2022stability,szolnoki2004phase,geiger2018topologically}.
Finally, higher-order interactions might be frequent in nature and affect resulting patterns \cite{palombi2020coevolutionary,gibbs2022coexistence,griffin2023higher}.

Beside these complex models, we also still lack basic understanding of (chemical) species that interact and react.
Along these lines, it will be interesting to investigate thermodynamic constraints on spatiotemporal patterns.
A recent manuscript already used linear stability analysis to investigate general non-ideal reaction-diffusion systems~\cite{Aslyamov2023}, and this work needs to be extended to include oscillating patterns.
It will be interesting to investigate fundamental physical constraints on creating spatiotemporal patterns, which will aid their reconstitution in experiments.

\tocless{\section*{Acknowledgements}}
CL thanks Lucas Menou for helping to set up the numerical calculations at the beginning of the project. We thank Yicheng Qiang for helpful discussions and critical reading of the manuscript.
We gratefully acknowledge funding from the Max Planck Society and the European Union (ERC, EmulSim, 101044662).

\onecolumngrid
\appendix
% \begin{appendix}
  % \label{appendix}
  \renewcommand{\thefigure}{S\arabic{figure}}
  \setcounter{figure}{0}

\section{Equilibrium model without chemical reactions}
For the model without chemical reactions, i.e., $R_i=0$, we apply linear stability analysis around $\phi_i=\phi_0$.
The three eigenvalues are 
\begin{align}
\lambda_1&=\lambda_2=D q^2 \left(-q^2 w^2 \phi_0+\chi  \phi_0-1\right)
& \text{and} &&
\lambda_3&=D q^2 \left(-q^2 w^2 \phi_0-2 \chi  \phi_0+\frac{1}{3 \phi_0-1}\right)
\;.
\end{align}
Consequently, the two spinodal curves for phase separation read
\begin{align}
  \chi_R&=\frac{1}{\phi_0}
& \text{and} &&
  \chi_A&=-\frac{1}{2\phi_0(1-3\phi_0)}
  \;.
\end{align}
When $\chi>\chi_R$, the highest growth rate is $\lambda_1=\lambda_2>0$; when $\chi<\chi_A$, the highest growth rate is $\lambda_3>0$. Taking $\phi_0=\frac14$, we get $\chi_R=4$ and $\chi_A=-8$. %

\section{Linear stability analysis of full model}
We here present details of the linear stability analysis of the dynamical equations, given by Eq. (3) in the main text.
We linearize  Eq. (3) in the main text around the uniform stationary state \begin{eqnarray}
  \phi_i=\phi_0=\beta/(3\beta+\sigma),
\end{eqnarray} and determine the time evolution of perturbations in Fourier space, where the perturbations are characterized by the wave vector $\vect q$. The stability is determined by the eigenvalues of the Jacobian matrix \cite{Cross2009}
\begin{eqnarray}
\vect J (\vect q)=\frac{1}{ 3 \beta +\sigma }(\vect C_0+\vect C_1\vect{q}^2+\vect C_2\vect{q}^4)\;,
\end{eqnarray}
where
\begin{eqnarray}
\vect C_0=\left(
\begin{array}{ccc}
 -\beta ^2-6 \beta  \mu -2 \mu  \sigma  & -\beta ^2+\beta  (\zeta +3 \mu )+\mu  \sigma  & -\beta ^2-\beta  (\zeta -3 \mu +\sigma )+\mu  \sigma  \\
 -\beta ^2-\beta  (\zeta -3 \mu +\sigma )+\mu  \sigma  & -\beta ^2-6 \beta  \mu -2 \mu  \sigma  & -\beta ^2+\beta  (\zeta +3 \mu )+\mu  \sigma  \\
 -\beta ^2+\beta  (\zeta +3 \mu )+\mu  \sigma  & -\beta ^2-\beta  (\zeta -3 \mu +\sigma )+\mu  \sigma  & -\beta ^2-6 \beta  \mu -2 \mu  \sigma  \\
\end{array}
\right),
\end{eqnarray}
\begin{eqnarray}
  \vect C_1=\frac{1}{\sigma}\left(
  \begin{array}{ccc}
   -\left(D (\beta +\sigma ) (3 \beta +\sigma )\right) & -\beta  D (3 \beta +\sigma  \chi +\sigma ) & -\beta  D (3 \beta +\sigma  \chi +\sigma ) \\
   -\beta  D (3 \beta +\sigma  \chi +\sigma ) & -\left((\beta +\sigma ) (3 \beta +\sigma ) D\right) & -\beta  D (3 \beta +\sigma  \chi +\sigma ) \\
   -\beta  D (3 \beta +\sigma  \chi +\sigma ) & -\beta  D (3 \beta +\sigma  \chi +\sigma ) & -\left((\beta +\sigma ) (3 \beta +\sigma ) D_C\right) \\
  \end{array}
  \right),
  \end{eqnarray}
and 
\begin{eqnarray}
  \vect C_2=\left(
  \begin{array}{ccc}
   -\beta  w^2D & 0 & 0 \\
   0 & -\beta   w^2D & 0 \\
   0 & 0 & -\beta   w^2D \\
  \end{array}
  \right).
  \end{eqnarray}
It can be seen that the interaction $\chi$ only appears in the off-diagonal elements in $\vect C_1$, which effectively only change the cross diffusion, similar to a system we have studied recently \cite{menou2023physical}. Comparing to the model in ref. \cite{szczesny2014coevolutionary}, where the Jacobian matrix is $\vect J_t=\frac{1}{ 3 \beta +\sigma }(\vect C_{0}+\vect C_{1t}\vect{q}^2)$ with
\begin{eqnarray}
\vect C_{1t}=\left(
\begin{array}{ccc}
 -\beta  (\delta_D+2 \delta_E)-\delta_D \sigma  & \beta  (\delta_E-\delta_D) & \beta  (\delta_E-\delta_D) \\
 \beta  (\delta_E-\delta_D) & -\beta  (\delta_D+2 \delta_E)-\delta_D \sigma  & \beta  (\delta_E-\delta_D) \\
 \beta  (\delta_E-\delta_D) & \beta  (\delta_E-\delta_D) & -\beta  (\delta_D+2 \delta_E)-\delta_D \sigma  \\
\end{array}
\right),
\end{eqnarray}
whose cross diffusion is from the difference of migration terms $\delta_E$ and $\delta_D$, in our model the cross diffusion appears naturally from the interactions.

The eigenvalues of $\vect J$ are  
\begin{subequations}
\begin{align}
\lambda_1&=\frac{\beta  (\sigma -18 \mu )-6 \mu  \sigma }{2 (3 \beta +\sigma )}-\frac{D  (\sigma -\beta  (\chi -3))}{3 \beta +\sigma }q^2-\frac{\beta  D }{3 \beta +\sigma }w^2q^4-i \frac{\sqrt{3} \beta  (2 \zeta +\sigma )}{2 (3 \beta +\sigma )}
\\
\lambda_2&=\frac{\beta  (\sigma -18 \mu )-6 \mu  \sigma }{2 (3 \beta +\sigma )}-\frac{D  (\sigma -\beta  (\chi -3))}{3 \beta +\sigma }q^2-\frac{\beta  D }{3 \beta +\sigma }w^2q^4+i \frac{\sqrt{3} \beta  (2 \zeta +\sigma )}{2 (3 \beta +\sigma )}
\\
\lambda_3&=-\beta -\frac{D \left(9 \beta ^2+2 \beta  \sigma  (\chi +3)+\sigma ^2\right)}{\sigma  (3 \beta +\sigma )} q^2 -\frac{\beta  D }{3 \beta +\sigma } w^2q^4
\;.
\end{align}
\end{subequations}
We denote $\lambda_r=\lambda_3$ and $\lambda_c=\lambda_2$, i.e., $\mathrm{Re}(\lambda_c)=\mathrm{Re}(\lambda_1)=\mathrm{Re}(\lambda_2)$ and $\mathrm{Im}(\lambda_c)=-\mathrm{Im}(\lambda_1)=\mathrm{Im}(\lambda_2)$ in the main text.

Let us first focus on $\lambda_r$. By solving 
\begin{eqnarray}
  \frac{\diff \lambda_r}{\diff (q_r^m)^2}=0
\end{eqnarray} 
and 
\begin{eqnarray}
  \lambda_r(q=q_r^m,\chi=\chi_A)=0
\end{eqnarray} 
we obtain the peak position of $\lambda_r$,
\begin{eqnarray}
  q^m_{r}=\sqrt{\frac{1}{w^2}\left(-\frac{(3 \beta +\sigma )^2}{2 \beta  \sigma }-\chi\right)}\;,
\end{eqnarray}
and the critical physical interaction
\begin{eqnarray}
\chi_A=-3-\frac{9 \beta }{2 \sigma }-\frac{\sigma }{2 \beta }-w\sqrt{\frac{3 \beta +\sigma }{D}}\;.
\end{eqnarray}
For $\chi>\chi_A$, $\mathrm{max}(\lambda_r)<0$ so the mode is stable for all wave vectors. In contrast, for $\chi<\chi_A$, $\mathrm{max}(\lambda_r)>0$. By solving 
\begin{eqnarray}
  \lambda_r(q=q_r^{\pm})=0
\end{eqnarray} 
we get the two zero points
\begin{eqnarray}
  q_r^{\pm}=\sqrt{\frac{-\left(2 \beta  \sigma  \chi +(3 \beta +\sigma )^2\right)\pm\sqrt{ \left(2 \beta  \sigma  \chi +(3 \beta +\sigma )^2\right)^2-4 \beta ^2 \sigma ^2 w^2 (3 \beta +\sigma )/D}}{2 \beta   \sigma  w^2}}\;.
\end{eqnarray}
Therefore, we obtain the unstable modes within the band $q_r^-<q<q_r^+$ for $\chi<\chi_A$. The critical $\chi=\chi_A$ corresponds to the blue curve in Fig. 1(b) in the main text.

Next we pay attention to $\lambda_c$. 
For $\mu>\mu_*$, we can see $\mathrm{Re}(\lambda_c) (q=0)<0$. We solve 
\begin{eqnarray}
  \frac{\diff \lambda_c}{\diff (q_c^m)^2}=0
\end{eqnarray} 
to obtain the peak position of $\mathrm{Re}(\lambda_c)$, which is 
\begin{eqnarray}
  q_c^m=\sqrt{\frac{  \chi -(3 +\sigma/\beta) }{2  w^2}}\;.
\end{eqnarray}
We then calculate the critical physical interaction $\chi_R$ by solving 
\begin{eqnarray}
  \mathrm{Re}(\lambda_c(q=q_c^m))=0\;,
\end{eqnarray}
which gives
\begin{eqnarray}
  \chi_R=3+\sigma/\beta+w\sqrt{2\sigma (\mu/\mu_*-1)/ D}\;.
\end{eqnarray}
Therefore, for $\chi<\chi_R$ and $\mu>\mu_*$, $\mathrm{Re}(\lambda_c)<0$ for any wave vector, and hence the mode is always stable. However, for $\chi>\chi_R$ and $\mu>\mu_*$,  $\mathrm{max}(\mathrm{Re}(\lambda_c))>0$. Solving 
\begin{eqnarray}
  \mathrm{Re}(\lambda_c(q=q_c^\pm))=0
\end{eqnarray}
gives the zeros points,
\begin{eqnarray}
  \label{eq:qcpm}
  q_c^\pm= \sqrt{\frac{  \tilde{\chi} \pm \sqrt{   \tilde{\chi} ^2-2  \sigma  w^2(\mu/\mu_*-1)/D}}{2   }}\;,
\end{eqnarray}
where $\tilde{\chi}=\chi-3-\sigma/\beta$.
Therefore, the modes $\lambda_c$ in the band $q_c^-<q<q_c^+$ are unstable. 

For $\mu<\mu_*$, we can see $\mathrm{Re}(\lambda_c) (q=0)>0$, which means the modes $\lambda_c$ is always unstable at $q=0$. However, we can still find a critical physical interaction by solving $q_c^m(\chi=\chi_R^*)=0$ whose solution is 
\begin{eqnarray}
  \chi_R^*=3+\frac{\sigma}{\beta}\;.
\end{eqnarray}
For $\chi<\chi_R^*$, $q_c^m=0$ and hence the unstable band becomes $0=q_c^m<q<q_c^+$; while for $\chi>\chi_R^*$, $q_c^m>0$ and the unstable band is within $q_c^-<q<q_c^+$ given by \Eqref{eq:qcpm}.

Meanwhile, we point out that the frequency from the imaginary part of the eigenvalues is
\begin{eqnarray}
\omega_*=\frac{\sqrt{3} \beta  (2 \zeta +\sigma )}{2 (3 \beta +\sigma )}\;.
\end{eqnarray}

Combining the critical physical interactions and the Hopf bifurcation $\mu=\mu_*$, we can separation the parameters space $(\chi,\mu)$ to six regimes, named AH, WH, RH, AL, WL, RL, as introduced in the main text.
The analysis above predicts the stability diagram, the eigenvalues and the length scales shown in Fig. 1 in the main text.

\section{Multiscale expansion of the full model}
This section uses the multiscale expansion presented in ref. \cite{szczesny2014coevolutionary} to characterize the dynamical behavior near the the Hopf bifurcation (for $\mu\lessapprox\mu_*$) when the length scale of patterns is large compared to $w$ and the physical interaction $\chi$ is weak.
We first perform a linear transformation of the fields to simplify the equations, then use the multiscale expansion to obtain the complex Ginzburg-Landau equation (CGLE), and finally map it back to the original model to obtain the scaling laws for the wavelength and frequency shown in the main text. %
\subsection{Linear transformation}
To further study the behavior of the model, we first use a linear transformation to simplify the system. We can shift the fields by $\phi_0$, introducing new fields $v_i=\phi_i-\phi_0$ for $i=A,\ B,\ C$. The linear equation 
\begin{eqnarray}
  \frac{\diff \vect{v}}{\diff t} = \vect J_0 \vect{v} + \mathcal{O}(\vect v)^2
\end{eqnarray}
with $\vect J_0\equiv\vect J(\vect q=0)=\frac{1}{3\beta+\sigma}\vect C_0$
can be transformed to a Jordan normal form
\begin{eqnarray}
  \frac{\diff \vect{u}}{\diff t}= \vect S^{-1}\vect J_0\vect S \vect{u}=\mathcal{J}_0\vect{u},
\end{eqnarray}
where we define the field $\vect{u}=\vect S^{-1}\vect{v}$. %
In the above equation, we also defined the matrices
\begin{align}
  \vect S&=\left(
    \begin{array}{ccc}
     -\frac{1}{\sqrt{6}} & -\frac{1}{\sqrt{2}} & \frac{1}{\sqrt{3}} \\
     -\frac{1}{\sqrt{6}} & \frac{1}{\sqrt{2}} & \frac{1}{\sqrt{3}} \\
     \sqrt{\frac{2}{3}} & 0 & \frac{1}{\sqrt{3}} \\
    \end{array}
    \right) \;,
&
\vect S^{-1}&=
\left(
\begin{array}{ccc}
 -\frac{1}{\sqrt{6}} & -\frac{1}{\sqrt{6}} & \sqrt{\frac{2}{3}} \\
 -\frac{1}{\sqrt{2}} & \frac{1}{\sqrt{2}} & 0 \\
 \frac{1}{\sqrt{3}} & \frac{1}{\sqrt{3}} & \frac{1}{\sqrt{3}} \\
\end{array}
\right)
&\text{and} &&
  \mathcal{J}_0&=\left(
    \begin{array}{ccc}
     \alpha  & -\omega_* & 0 \\
     \omega_* & \alpha  & 0 \\
     0 & 0 & -\beta  \\
    \end{array}
    \right)
    \;,
\end{align}
where $\alpha=\frac{\beta  (\sigma -18 \mu )-6 \mu  \sigma }{2 (3 \beta +\sigma )}$.
Applying the above transformation to the full equation Eq. 3 in the main text, we obtain
\begin{eqnarray}
	\label{eqn:pde_linear_transform}
  \frac{\diff \vect{u}}{\diff t}=S^{-1}\frac{\diff \vect{v}}{\diff t} = S^{-1}\frac{\diff \vect{\phi}}{\diff t} 
  \;,
\end{eqnarray}
where the time derivative of $\vect{\phi} = \{\phi_A, \phi_B, \phi_C\}$ is defined in Eq. 7 in the main text.
Replacing the fields~$\vect{\phi}$ by the transformed fields $\vect u$ results in a new set of partial differential equations, which we analyze in the following.

\subsection{Multiscale expansion approach}
We next perform a multiscale expansion around the Hopf bifurcation to obtain the complex Ginzburg Landau equation (CGLE) following \cite{szczesny2014coevolutionary}.
A space and time perturbation expansion in the parameter $\epsilon = \sqrt{3(\mu_*-\mu)}$ is performed by introducing the `slow variables' 
\begin{eqnarray}
  (T,X)=(\epsilon^2 t,\epsilon x)
\end{eqnarray} and expanding the transformed densities $\vect u$ in powers of $\epsilon$. Here we use $x$ to represent spatial coordinates $\br$ for simplicity. More specifically, we use the multiscale expansion of time and space coordinates for an arbitrary function $F(t,x,T,X)$,
\begin{subequations}
\begin{align}
  \frac{\diff F(t,x,T,X)}{\diff t}
  	&=\frac{\partial F(t,x,T,X)}{\partial t}+\frac{\partial F(t,x,T,X)}{\partial T}\frac{\partial T}{\partial t}=\frac{\partial F(t,x,T,X)}{\partial t}+\epsilon^2\frac{\partial F(t,x,T,X)}{\partial T}
\\
  \frac{\diff F(t,x,T,X)}{\diff x} &=\frac{\partial F(t,x,T,X)}{\partial x}+\frac{\partial F(t,x,T,X)}{\partial X}\frac{\partial X}{\partial x}=\frac{\partial F(t,x,T,X)}{\partial x}+\epsilon\frac{\partial F(t,x,T,X)}{\partial X}\;,
\end{align} %
\end{subequations}
which implies
\begin{align}
  \frac{\partial}{\partial t}&\rightarrow\frac{\partial}{\partial t}+\epsilon^2\frac{\partial}{\partial T}
& \text{and} &&
  \frac{\partial}{\partial x}&\rightarrow\frac{\partial}{\partial x}+\epsilon\frac{\partial}{\partial X}
  \;.
\end{align}
We then start from \Eqref{eqn:pde_linear_transform} 
and assume $u_i(x,t)=\sum_{n=1}^{3} \epsilon^n U_{i,n}(t,T,X)$.
To order $\epsilon$, we then have
\begin{eqnarray}
  \left(
      \begin{array}{c}
   \frac{ \partial U_{1,1}(t,T,X)}{\partial t}\\
   \frac{ \partial U_{2,1}(t,T,X)}{\partial t}\\
   \frac{ \partial U_{3,1}(t,T,X)}{\partial t}
  \end{array}
  \right)
  =\left(
    \begin{array}{c}
     -\omega _* U_{2,1}(t,T,X) \\
     \omega _* U_{1,1}(t,T,X) \\
     -\beta  U_{3,1}(t,T,X) \\
    \end{array}
    \right)
\end{eqnarray}
Defining $Z_n=U_{1,n}+iU_{2,n}$, the equation simplifies to 
 \begin{eqnarray}
  \left(
    \begin{array}{c}
      \frac{\partial Z_1(t,T,X)}{\partial t}\\
      \frac{ \partial U_{3,1}(t,T,X)}{\partial t}
    \end{array}
    \right)=\left(
      \begin{array}{c}
        i \omega _* Z_1(t,T,X)  \\
        -\beta  U_{3,1}(t,T,X)\\
      \end{array}
      \right).
\end{eqnarray}
We next use the Ansatze  
\begin{equation}
  \label{eq:Z1}
  Z_1(t,T,X)=A_1(T,X)e^{i\omega_* t}
  \;,
\end{equation}
with $A_1(T,X)$ to be determined later
and $U_3^{(1)}=0$.
At the second order of $\epsilon$, we find
\begin{eqnarray}
  \left(
    \begin{array}{c}
      \frac{\partial Z_2(t,T,X)}{\partial t}
      \\
      \frac{ \partial U_{3,2}(t,T,X)}{\partial t}
      \end{array}
    \right)=
    \left(
      \begin{array}{c}
        \frac{e^{-2 i t \omega _*} \left(\beta  \sigma -2 i (3 \beta +\sigma ) \omega _*\right) \left(A_1(T,X){}^*\right){}^2}{2 \sqrt{6} \beta }+i \omega _* Z_2(t,T,X)
        \\
        \frac{1}{6} \left(\sqrt{3} \sigma  A_1(T,X) A_1(T,X){}^*-6 \beta  U_{3,2}(t,T,X)\right)
  \end{array}
    \right)
  \;,
\end{eqnarray}
where the superscript star $^*$ represents the complex conjugate operation.
The second equation gives $U_{3,2}=\frac{\sqrt{3}}{6}\frac{\sigma}{\beta}|Z^{(1)}|^2$. %
We propose the Ansatze of $Z_2(t,T,X)$ from the first equation,
\begin{eqnarray}
  Z_2(t,T,X)=
  \frac{i e^{-2 i \omega _* t} \left(\beta  \sigma -2 i (3 \beta +\sigma ) \omega _*\right) \left(A_1(T,X){}^*\right){}^2}{6 \sqrt{6} \beta  \omega _*}
  +A_2(T,X) \exp \left(i \omega _* t \right)
  \;,
\end{eqnarray}
with $A_2(T,X)$ to be determined. 
At the third order of $\epsilon$, we obtain 
\begin{align}
  e^{i t \omega _H} &\frac{\partial A_1(T,X)}{\partial T}+\frac{\partial Z_3(t,T,X)}{\partial t}
  \notag\\
  &=
  e^{i t \omega _H}\bigg[A_1(T,X)+\frac{-i \left(\beta ^2 \sigma ^2-3 i \beta  \sigma  (6 \beta +\sigma ) \omega _H+2 (6 \beta -\sigma ) (3 \beta +\sigma ) \omega _H^2\right)  }{36 \beta ^2 \omega _H}\left| A_1(T,X)\right| {}^2A_1(T,X)
  \notag  \\
  &\qquad\qquad+\frac{D (-\beta  \chi +3 \beta +\sigma )  \nabla^2_X A_1(T,X)}{3 \beta +\sigma }\bigg]
  \notag  \\
  &\quad
  +i \omega _H Z_3(t,T,X)+\frac{e^{-2 i t \omega _H} \left(\beta  \sigma -2 i (3 \beta +\sigma ) \omega _H\right) A_1(T,X){}^* A_2(T,X){}^*}{\sqrt{6} \beta }
  \;.
\end{align}
To remove the secular term, i.e., the term proportional to $e^{i t \omega _H}$,  we obtain
\begin{eqnarray}
  \partial A_1 /\partial T=\delta \nabla_X^2 A_1 +A_1 -(c_r+ic_i)|A_1|^2A_1,
\end{eqnarray}
where
\begin{align}
  c_r&=\frac{\sigma  (6 \beta +\sigma )}{12 \beta } \;,
&
  c_i&=\frac{1}{36} \left(\frac{2 (6 \beta -\sigma ) (3 \beta +\sigma ) \omega _*}{\beta ^2}+\frac{\sigma ^2}{\omega _*}\right)\;,
& \text{and} &&
  \delta&=D-\frac{\beta  D \chi }{3 \beta +\sigma }\;.
\end{align}
Note both $c_r$ and $c_i$ are the same as in \cite{szczesny2014coevolutionary}. 
Moreover, if we replace 
\begin{align}
  \delta_D &= D \left(\frac{2 \beta  \chi }{3 \beta +\sigma }+\frac{3 \beta }{\sigma }+1\right)
  & \text{and} &&
  \delta_E&= -\frac{D \chi  (\beta +\sigma )}{3 \beta +\sigma }
\end{align}
or
\begin{align}
  D&= \frac{\sigma  (\beta  (\delta_D+2 \delta_E)+\delta_D \sigma )}{(\beta +\sigma ) (3 \beta +\sigma )}
&\text{and}&&
  \chi &= -\frac{\delta_E (3 \beta +\sigma )^2}{\sigma  (\beta  (\delta_D+2 \delta_E)+\delta_D \sigma )}
  \;,
\end{align}
the $\delta$ in Eq.~(2.68) in ref. \cite{szczesny2014coevolutionary} is identical to our result.

If we use $A=\sqrt{c_r}A_1$, the equation can be further simplified to  
\begin{eqnarray}
  \frac{\partial A}{\partial T}=\delta \nabla_X^2 A +A -(1+ic)|A|^2A\;,
\end{eqnarray}
where $c=c_i/c_r$ reads
\begin{eqnarray}
  \label{eq:c_zeta}
  c=\frac{12\zeta(6\beta-\sigma)(\sigma+\zeta)+\sigma^2(24\beta-\sigma)}{3\sqrt{3}\sigma(6\beta+\sigma)(\sigma+2\zeta)}\;.
\end{eqnarray}
Note that $\delta$ can further be absorbed into the length scale. That is, if we define $X'=X/\sqrt{|\delta|}$, the equation becomes
\begin{eqnarray}
  \label{eq:CGLE}
  \partial A /\partial T=\sgn(\delta) \nabla_{X'}^2 A+A -(1+ic)|A|^2 A.
\end{eqnarray}
Taken together, near the Hopf bifurcation, the effect of $\chi$ is similar to the nonlinear diffusivity and hence we can also get four phases~\cite{szczesny2013does,szczesny2014characterization}: absolute instability (AI), Eckhaus instability (EI),
bound states (BS) and spiral annihilation (SA).

\subsection{Mapping to complex Ginzburg-Landau equation and scaling laws}
Denoting $X=X'$ for simplicity, let us first study the plane wave solution of \Eqref{eq:CGLE}
\begin{align}
  A&=Ue^{i(k_{\CGLE}X-\Omega_{\CGLE} T)}
& \text{with} &&
  \Omega_{\CGLE}&=c|U|^2=c(1-\sgn(\delta) k_{\CGLE}^2)\;.
\end{align}
Therefore, we obtain the wavelength $\lambda_{\CGLE}=2\pi/k_{\CGLE}$,
\begin{eqnarray}
  \label{eq:lCGLE}
  \lambda_{\CGLE}=2 \pi \sqrt{\frac{\sgn(\delta)}{1-|U|^2}} \;,
\end{eqnarray}
and velocity $v_{\CGLE}=\Omega_{\CGLE}/k$,
\begin{eqnarray}
  v_{\CGLE}=c|U|^2\sqrt{\frac{\sgn(\delta)}{1-|U|^2}}\;,
\end{eqnarray}
as a function of the amplitude~$U$.

To map to the full model, we make use of $X=\frac{\epsilon}{\sqrt{|\delta|}} x$ and $T=\epsilon^2 t$, where $\epsilon=\sqrt{3(\mu_*-\mu)}$. We obtain the wavelength 
\begin{eqnarray}
  \label{eq:lCGLE2}
  \lambda_{\CGLE}%
  =\sqrt{\frac{3(\mu_*-\mu)}{D|1-\frac{\beta\chi}{3\beta+\sigma}|}}\lambda\;.
\end{eqnarray}
Combining Eqs.~\eqref{eq:lCGLE} and \eqref{eq:lCGLE2}, we can predict the wavelength from the amplitude $|U|^2$ of the CGLE,
\begin{eqnarray}
  \lambda=2 \pi \sqrt{\frac{D(1-\frac{\beta\chi}{3\beta+\sigma})}{3(\mu_*-\mu)(1-|U|^2)}}\;.
\end{eqnarray}

To compare to the frequency $\omega$ obtained from the correlation of $\phi_i(t)$, we substitute the solution of $A(X,T)$ into $Z_1(t,T,X)$ (\Eqref{eq:Z1}), %
\begin{eqnarray}
  Z_1(t,T,X)\sim e^{i(\omega_* t- \Omega_{\CGLE}T)}=e^{i(\omega_* - 3(\mu_*-\mu)\Omega_{\CGLE})t}=e^{i(\omega_* - 3(\mu_*-\mu)c|U|^2)t} \;.
\end{eqnarray}
We thus obtain the frequency that is comparable to that from simulation, i.e.
\begin{eqnarray}
  \omega=\omega_* - 3(\mu_*-\mu)c|U|^2\;.
\end{eqnarray}

\subsection{Numerical solution of the CGLE}
\begin{figure}
  \centering
  \includegraphics[width=1.0\columnwidth]{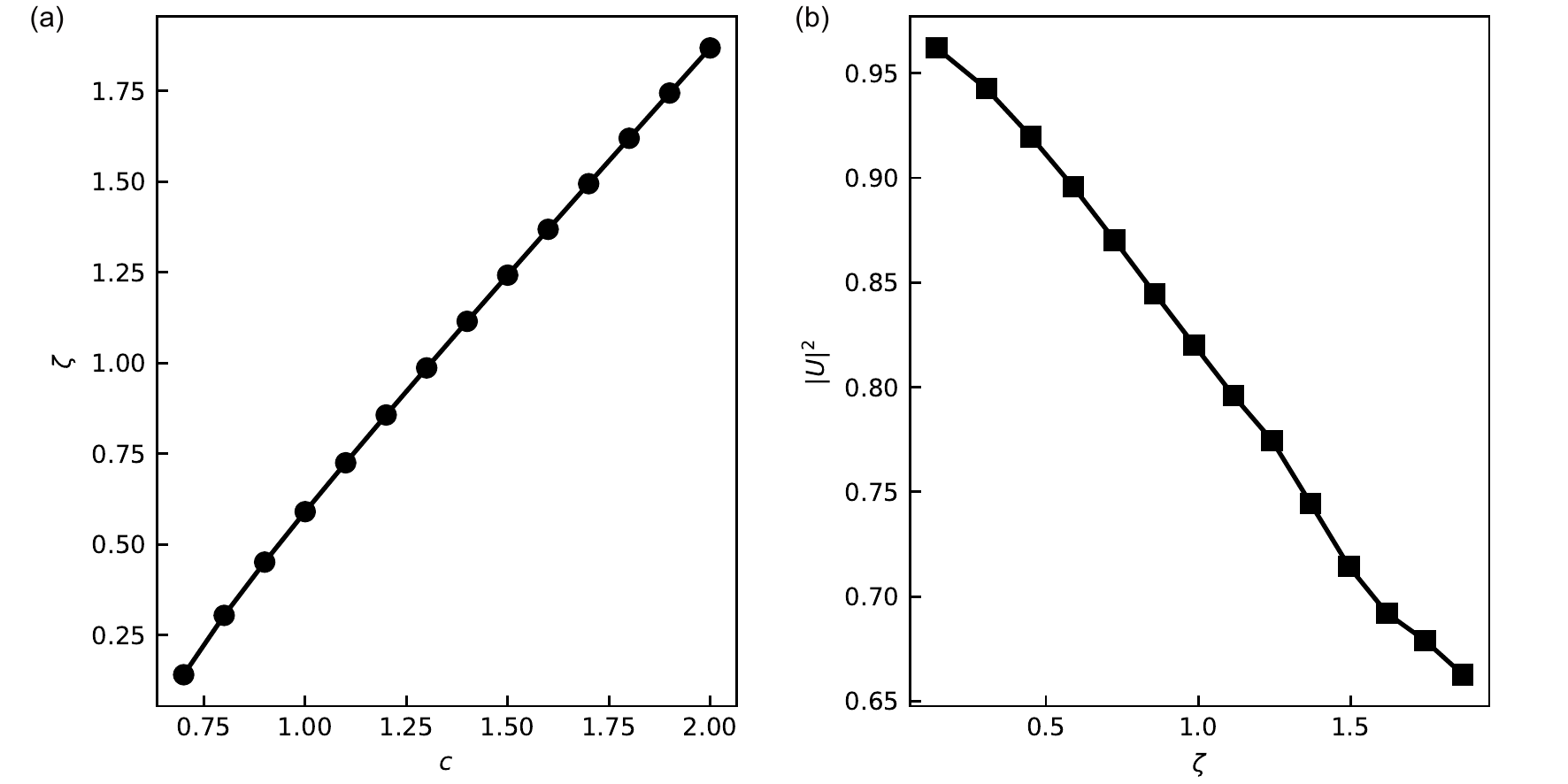}
  \caption{Numerical CGLE solutions. (a) Theoretical relation of $\zeta$ and $c$ using \Eqref{eq:c_zeta}. (b) Numerical results of $|U|^2$ as a function of $\zeta$ shown in (a). }
  \label{fig:S_CGLE} 
\end{figure}
We perform numerical simulation of the CGLE given by \Eqref{eq:CGLE} using $\sgn (\delta)=1$. \figref{fig:S_CGLE} shows the relation of $\zeta$ and $c$ and the relation of $\zeta$ and $|U|^2$.  The data are consistent with the results in \cite{szczesny2014coevolutionary}.

\section{Additional simulation snapshots}
\begin{figure}
  \centering
  \includegraphics[width=1.0\columnwidth]{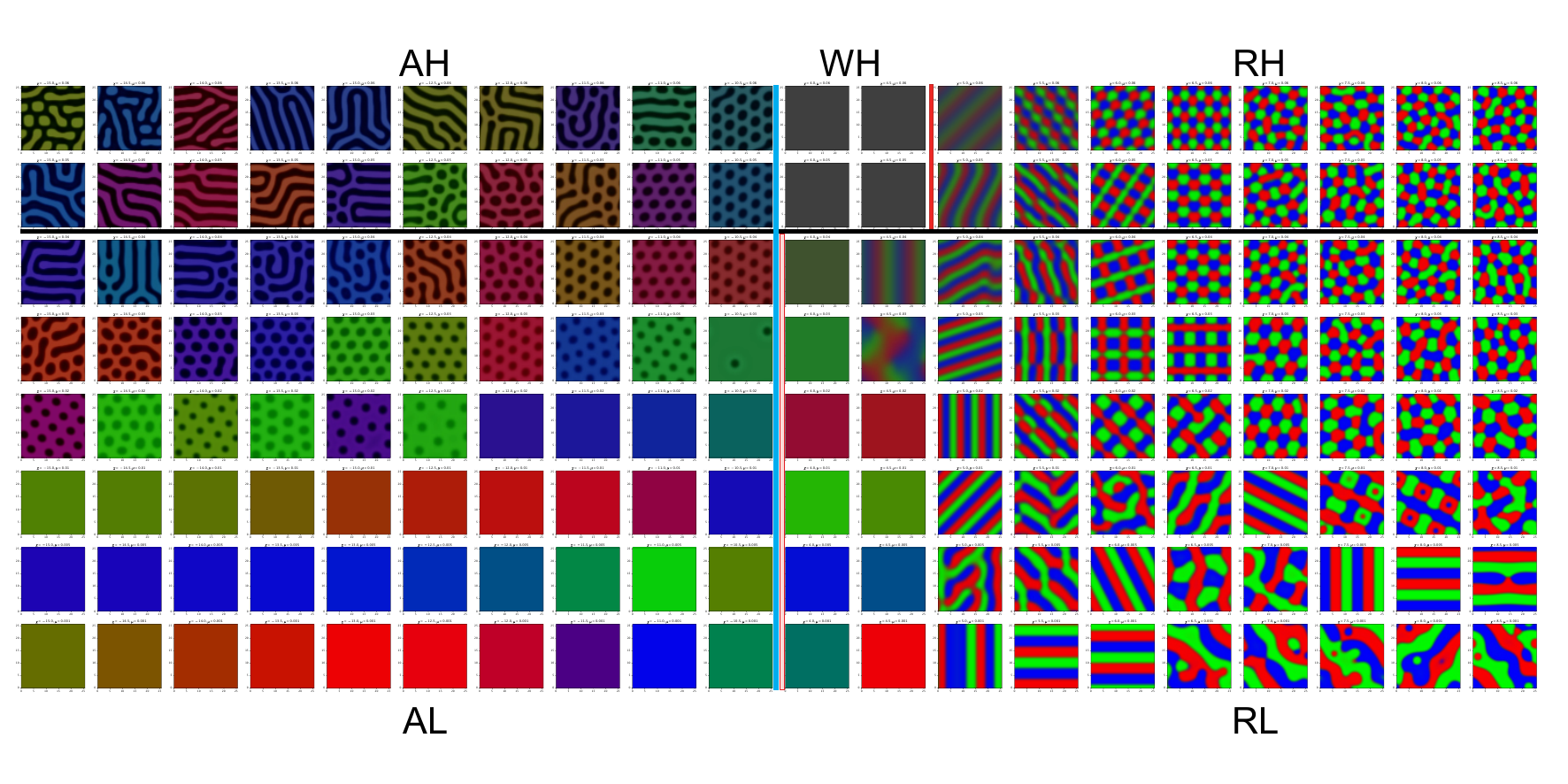}
  \caption{Snapshots from simulations at time $t=10^5$ for various~$\chi$ and $\mu$ using $D=\beta=\alpha=1$ and $\zeta=0.6$, using $L_x/w=L_y/w=25.6$, $\Delta x/w=\Delta y/w=0.4$.}
  \label{fig:S_snapshots_size64} 
\end{figure}

\begin{figure}
  \centering
  \includegraphics[width=1.0\columnwidth]{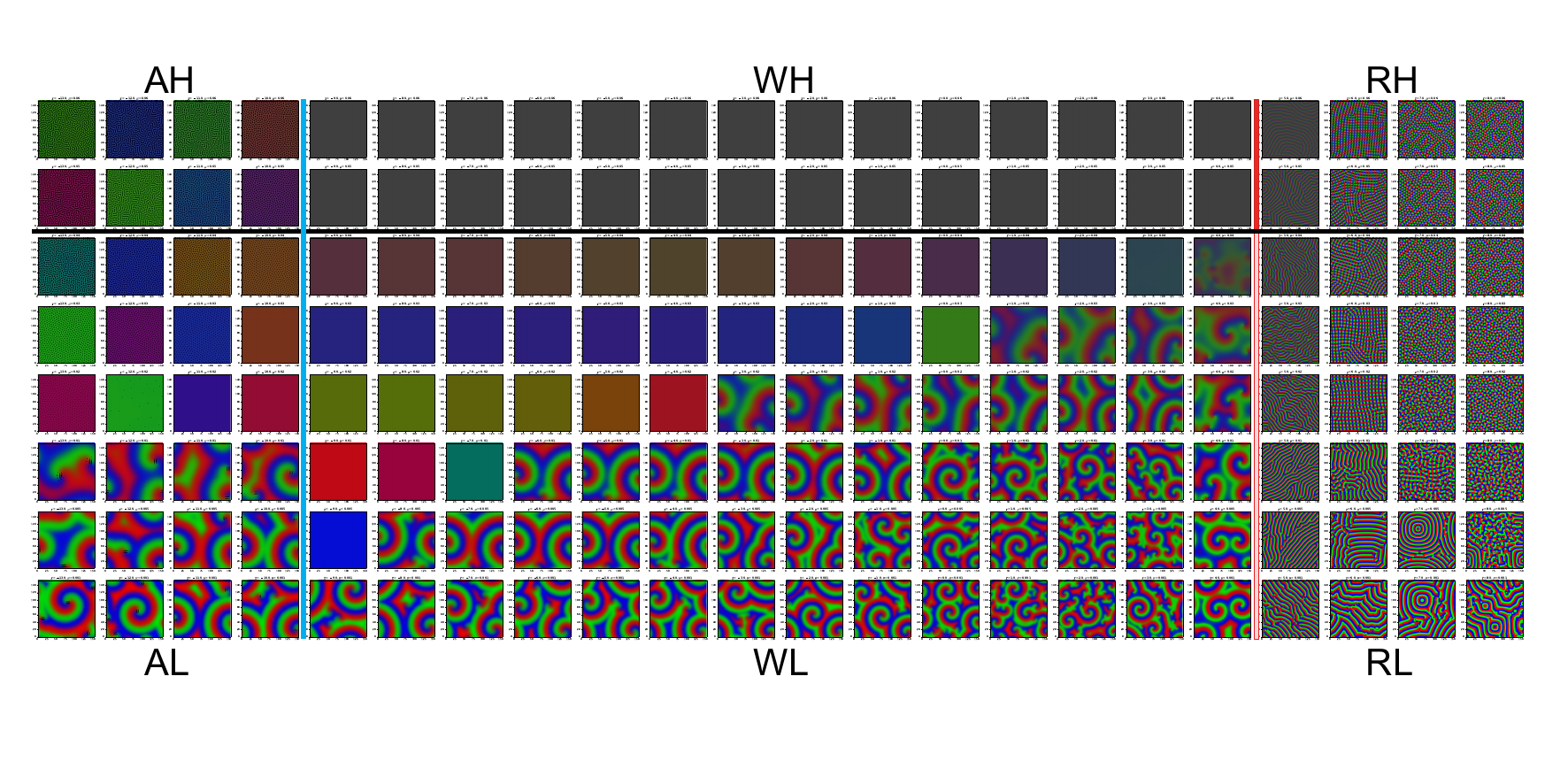}
  \caption{Snapshots from simulations at time $t=10^4$ for various~$\chi$ and $\mu$ using $D=\beta=\alpha=1$ and $\zeta=0.6$, using $L_x/w=L_y/w=153.6$, $\Delta x/w=\Delta y/w=0.6$.}
  \label{fig:S_snapshots_size256} 
\end{figure}

\begin{figure}
  \centering
  \includegraphics[width=1.0\columnwidth]{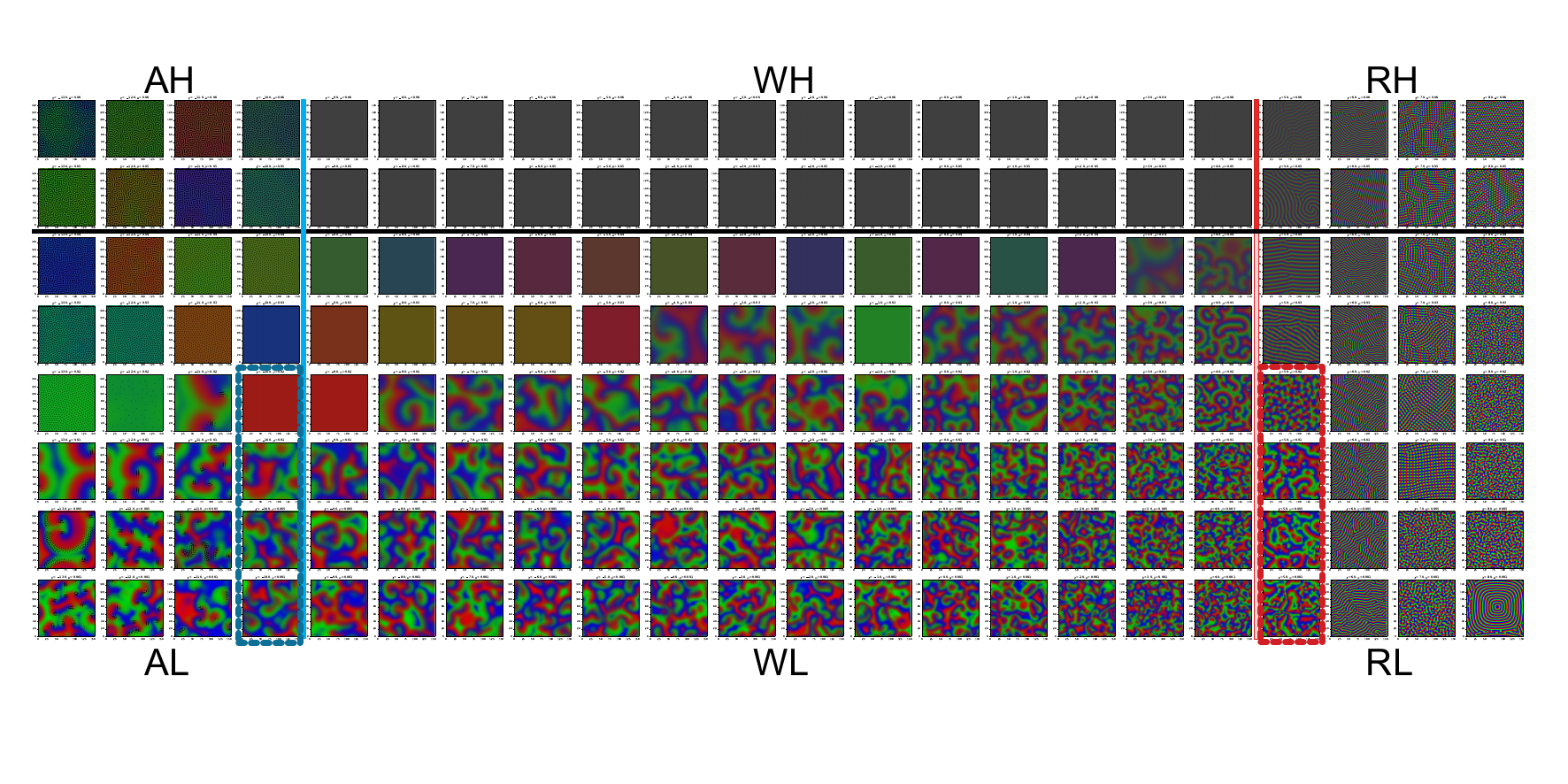}
  \caption{Snapshots from simulations at time $t=10^4$ for various~$\chi$ and $\mu$ using $D=\beta=\alpha=1$ and $\zeta=1.8$, using $L_x/w=L_y/w=153.6$, $\Delta x/w=\Delta y/w=0.6$.}
  \label{fig:S_snapshots_zeta18} 
\end{figure}

We show the simulation snapshots for $\zeta=0.6$ at different $\mu$ and $\chi$ in \figref{fig:S_snapshots_size64} and \figref{fig:S_snapshots_size256}. In \figref{fig:S_snapshots_zeta18} we show the snapshots for $\zeta=1.8$ and it can be seen that at such high replacement rate, the spiral-wave-like patterns in WL extend to $\chi<\chi_\mathrm{A}$ and $\chi>\chi_\mathrm{R}^*$, as labeled by the blue box and the red box.


% \end{appendix}
  
\clearpage

\bibliographystyle{apsrev4-1}
\bibliography{main}
\end{document}